\documentclass[a4paper,conference]{IEEEtran}

\usepackage{ifpdf}
\usepackage{cite}
\usepackage[pdftex]{graphicx}
\usepackage{array}
\usepackage{mdwmath}
\usepackage{mdwtab}
\usepackage{amssymb,latexsym}
\usepackage{stfloats}
\usepackage{amsmath}
\usepackage{subfig}
\usepackage{hyperref}

\hypersetup{
    colorlinks=true,
    }

\usepackage{algcompatible}

\usepackage{algorithm}

\usepackage[font={footnotesize}]{caption}

\usepackage{balance}

\usepackage[utf8]{inputenc}

\usepackage[T1]{fontenc}

\usepackage{algorithm}
\usepackage{algpseudocode}

\renewcommand{\figurename}{Figure}

\DeclareRobustCommand*{\IEEEauthorrefmark}[1]{%
\raisebox{0pt}[0pt][0pt]{\textsuperscript{\footnotesize\ensuremath{#1}}}}

\renewcommand\IEEEkeywordsname{Keywords}

\makeatletter
\def\ps@IEEEtitlepagestyle{%
  \def\@oddfoot{\mycopyrightnotice}%
  \def\@evenfoot{}%
}
\def\mycopyrightnotice{%
  {\footnotesize\begin{tabular}[t]{@{}l@{}} © 2022 IEEE.  Personal use of this material is permitted.  Permission from IEEE must be obtained for all other uses, in any current or future media, including \\ reprinting/republishing this material for advertising or promotional purposes, creating new collective works, for resale or redistribution to servers or lists,\\ or reuse of any copyrighted component of this work in other works.\end{tabular}}
  \gdef\mycopyrightnotice{}
}

\begin{document}

\title{Using the Polar Transform for Efficient Deep Learning-Based Aorta Segmentation in CTA Images} 
\author{\IEEEauthorblockN{
Marin Benčević*\IEEEauthorrefmark{1},
Marija Habijan\IEEEauthorrefmark{1},
Irena Galić\IEEEauthorrefmark{1},
Danilo Babin\IEEEauthorrefmark{2}}
\IEEEauthorblockA{\IEEEauthorrefmark{1}
Faculty of Electrical Engineering, Computer Science and Information Technology\\
J. J. Strossmayer University Osijek, Croatia}
\IEEEauthorblockA{\IEEEauthorrefmark{2}
imec-TELIN-IPI, Faculty of Engineering and Architecture\\
Ghent University, Belgium}
{*\it marin.bencevic@ferit.hr}}

\maketitle

\begin{abstract}
Medical image segmentation often requires segmenting multiple elliptical objects on a single image. This includes, among other tasks, segmenting vessels such as the aorta in axial CTA slices. In this paper, we present a general approach to improving the semantic segmentation performance of neural networks in these tasks and validate our approach on the task of aorta segmentation. We use a cascade of two neural networks, where one performs a rough segmentation based on the U-Net architecture and the other performs the final segmentation on polar image transformations of the input. Connected component analysis of the rough segmentation is used to construct the polar transformations, and predictions on multiple transformations of the same image are fused using hysteresis thresholding. We show that this method improves aorta segmentation performance without requiring complex neural network architectures. In addition, we show that our approach improves robustness and pixel-level recall while achieving segmentation performance in line with the state of the art.
\end{abstract}

\begin{IEEEkeywords}
Convolutional neural network; medical image processing; medical image segmentation; semantic segmentation.
\end{IEEEkeywords}

\IEEEpeerreviewmaketitle

\section{Introduction}
\label{intro}

The aorta is the largest artery of the human body and supplies oxygenated blood from the heart to all parts of the body. It is one of the most clinically significant structures to analyze for cardiovascular disease diagnosis and prevention.

Several conditions could occur on the aorta which can be detected using 3D medical imaging, including aneurysms, dissections, stenoses, coarctations, or traumas. All of these conditions are potentially dangerous and require careful screening, following, and potentially surgical treatment, while a failure or delay in the diagnosis of these conditions could be fatal. Therefore, developing a fully automated method to efficiently and accurately segment the aorta could be beneficial for earlier detection of these conditions. By producing a 3D model of the aorta from CT or MRI scans, a computer algorithm could perform automatic measurements to screen and detect aortic aneurysms, dissections, and other conditions which are commonly diagnosed by imaging the aorta.

In this paper, we present a new method for segmenting the aorta using deep neural networks. We do this by combining two neural networks, where one network performs the initial segmentation on 2D slices, which are then used to preprocess the input images using the polar transformation to better segment each connected component in the slice. 

We show that this method is comparable to the state-of-the-art aorta segmentation methods while being robust to small dataset sizes. In addition, we extend the method presented in \cite{bencevicTrainingPolarImage2021} and further validate the use of polar transformations in neural networks for medical image segmentation. These modifications can be used to improve performance in a  wide variety of medical image processing tasks where multiple elliptical objects need to be segmented, and can be added to existing methods for 2D image semantic segmentation without changing the underlying architecture.

\subsection{Related work}\label{related}

The method presented in this paper is an extension and improvement over the method presented in \cite{bencevicTrainingPolarImage2021}. The method uses a neural network to predict the center point of a single object in an input image. This center point is used as the origin of a polar transformation of the original input image. The transformed image is then segmented using a second neural network trained on polar image transformations. This approach works best when there is a single object on the image, i.e. one connected component in the segmentation label. In this paper, we modify the method to add support for multiple connected components by transforming each object to polar coordinates separately and then fusing the segmentations.

Several methods based on deep learning were proposed for segmenting the aorta from CT images. Fantazzini \textit{et al.} \cite{fantazzini3DAutomaticSegmentation2020} use a cascade of U-Net-based networks. They first perform a rough segmentation on axial slices to extract a region of interest. They then use separate networks to segment axial, sagittal, and coronal slices of the region of interest. Several other papers have used 3D U-Net-based architectures for this task \cite{yuThreeDimensionalDeepConvolutional2021, chenMultistageLearningSegmentation2021}.

The use of polar image transforms in neural networks was explored previously, including in the field of medical image segmentation \cite{ liuDDNetCartesianpolarDualdomain2019a}. Esteves \textit{et al.} \cite{estevesPolarTransformerNetworks2018b} train an end-to-end network which predicts a polar origin, transforms the image to polar coordinates and then performs classification. While similar to our approach, it differs in the application (segmentation and not classification) as well as in the neural network architectures used since we use two separate networks, making our approach simpler to adapt to existing architectures. Salehinejad \textit{et al.} \cite{salehinejadImageAugmentationUsing2018b} use multiple polar transformations of the same image as a method of data augmentation. The final labels are obtained using majority voting from predictions of multiple transforms of the same input image. In contrast, we employ per-object weighting and hysteresis thresholding to obtain the final prediction.

\section{Methodology}

\begin{figure*}
\centering
\includegraphics[width=1.97\columnwidth]{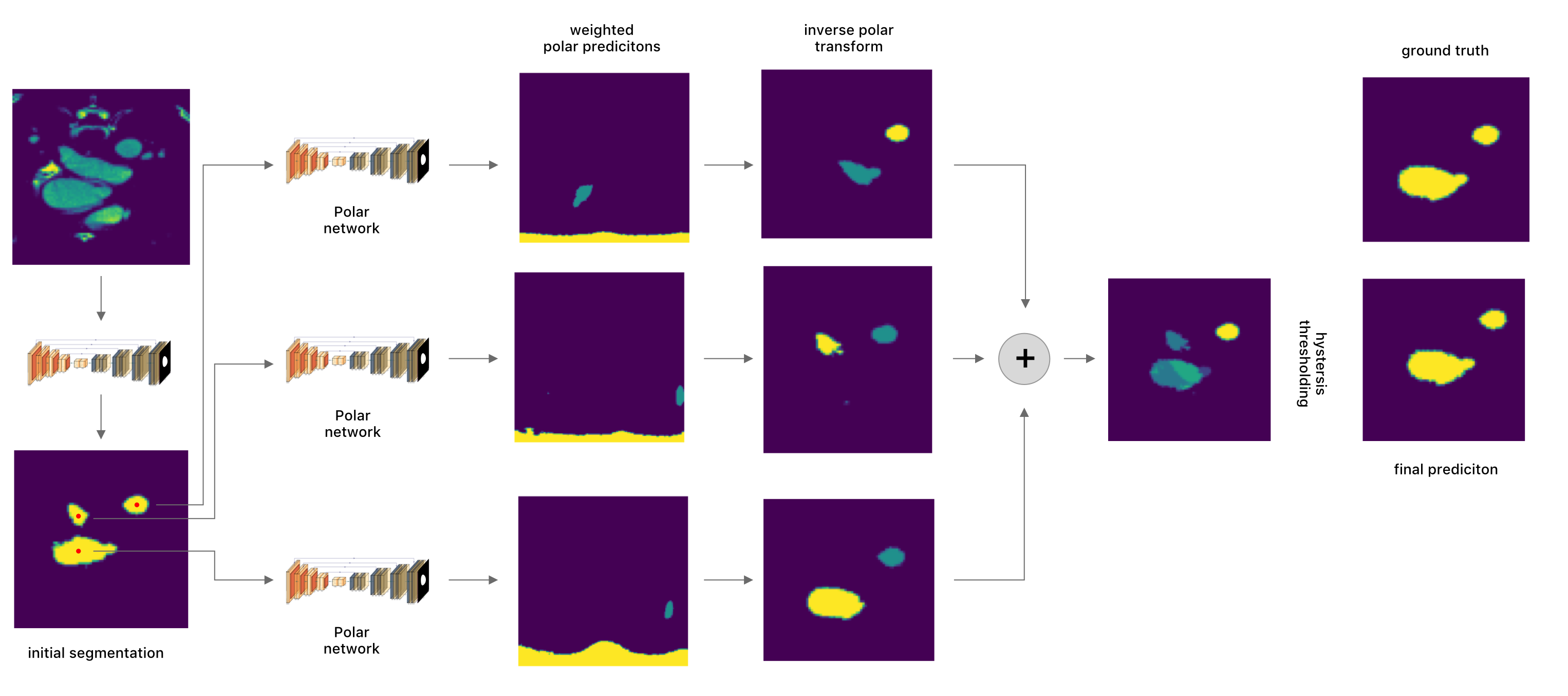}
\caption{A summary of our approach. An input image is first segmented using a U-Net network. For each connected component in the segmentation, the input image is transformed to polar coordinates using the centroid of the connected component as the origin. These images are then fed into a U-Net trained on polar images, and the predictions for each object are fused, hysteresis thresholded and transformed back to cartesian coordinates. Note how one of the false positive connected components in the initial segmentation was removed during hysteresis thresholding, since the component was only predicted in one of the three polar predictions.}
\label{fig:summary}
\end{figure*}

The work presented in this paper is an extension of \cite{bencevicTrainingPolarImage2021}. We obtain the center points of the objects in the image using a rough segmentation from a U-Net-based network, instead of using a center point predictor as is described in the original paper. We refer readers to the original paper for more details. A summary of our approach is shown in Figure \ref{fig:summary}.

In this paper, we perform several key modifications to allow the network to segment multiple objects on an image. During training of the polar network, we construct a dataset that contains one polar transformation per connected component in the ground truth segmentation label. The origins of these polar transformations are the centroid of each corresponding connected component. This has several advantages. First, one object is always the bottom part of the image to be segmented, so the network does not need to learn to first localize the object. Secondly, a consequence of this approach is that images with multiple connected components are over-sampled during training. This is a benefit since these images are both under-represented (when compared to images with a single connected component) and harder to segment (since they require segmenting multiple objects). In addition, during the training of the polar network, we use jittering of the center point when constructing the polar transformation to make the network more robust to inaccurate center point predictions. As an augmentation step, during each training step there is a $30\%$ chance that the polar origin will be shifted by a maximum of $\pm3$ pixels in any direction. This increases the robustness to inaccurate origin predictions during inference.

\begin{figure}
\centering
\includegraphics[width=\columnwidth]{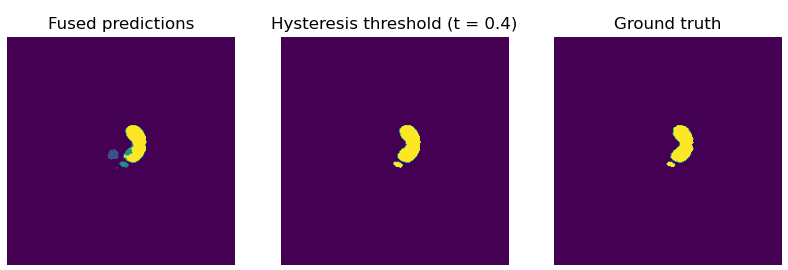}
\caption{Hystersis-thresholded segmentation output. For each polar prediction, the component which contains the origin of the transform gets a weight of 2 assigned, while all other components get a weight of 1. This left-most image is the result of summing the predictions of 3 polar transformations of the original image (one for each connected component), converted to cartesian coordinates. Note how the thresholding removes the false positive object on the left of the image while keeping the true positive objects intact.}
\label{fig:thresh}
\end{figure}

We also employ prediction fusion during inference. First, a 2D U-Net-based network is used to obtain an initial rough segmentation. A separate polar transformation for each connected component in the rough segmentation is constructed using the centroid of each component as the origin. The polar network predicts a segmentation map for each transform, resulting in a number of predictions equal to the number of connected components. In the predicted image, a weight of 2 is assigned to the connected component which contains the origin for that prediction, and a weight of 1 to all other connected components. We then sum all of the weighted images together. As shown in \cite{bencevicTrainingPolarImage2021}, the polar network generally performs best on objects which contain the polar origin, and worse at predicting other objects on the image. Therefore, we assign a larger weight to that component as a proxy for a confidence measure. We then sum all of the weighted predictions together and normalize the prediction to a 0-1 range. This leads to a segmentation map where each non-zero pixel represents the confidence that the pixel belongs to the aorta class. To obtain the final segmentation, we use hysteresis thresholding where the bottom threshold is 0, and the top threshold is 0.4, empirically determined according to the best Dice coefficient on the validation dataset. An example of thresholding a prediction is shown in Figure \ref{fig:thresh}.

\subsection{Data description and preprocessing}

We used a publicly available dataset of CT scans with corresponding aorta labels \cite{radlAVTMulticenterAortic2022} including the ascending aorta, the aortic arch as well as the descending and abdominal aorta. While the original dataset contains scans from three different centers, in our experiments we only use the data from Dongyang Hospital. In total, we use 18 CT scans, each containing 122-251 slices, with a slice thickness of 2 or 3 mm.

Each CT slice is windowed to a range of 200 to 500 HU to remove information from irrelevant tissues, then normalized to a range of -0.5 to 0.5, and zero centered using the global mean value across all slices in the validation set. The slices were each resized from $512 \times 666$ to $256 \times 256$ pixels. We use augmentation during training for both the cartesian and the polar network. The augmentations we use include a $50\%$ chance of a random combination of affine transforms including a shift of up to $6.25\%$, a scale of up to $10\%$ and a rotation of up to $15^{\circ}$; as well as a $30\%$ chance of a horizontal flip.

\subsection{Implementation details}

All of our models were implemented using PyTorch 3.9 using an NVIDIA GeForce RTX 3080 GPU. We use the U-Net \cite{ronnebergerUNetConvolutionalNetworks2015b} architecture for both the cartesian and the polar network. For training, we use a batch size of 8 and the Adam optimizer with a learning rate of $0.001$. All models were trained for 60 epochs with checkpointing where the model with the best validation Dice coefficient was selected. We use the Dice loss function as described in \cite{bencevicTrainingPolarImage2021}. All of the code, as well as the trained networks, can be found at \href{https://github.com/marinbenc/medical-polar-training}{github.com/marinbenc/medical-polar-training}

\section{Results and discussion}

\begin{table*}
\def\arraystretch{1.25}
\centering
\begin{tabular}{l c c c c} 
 \hline
 Method & DSC & mIoU & Prec. & Rec. \\ 
 \hline
U-Net (non-polar) & $0.886 \pm 0.049$ & $0.825 \pm 0.052$ & $0.901 \pm 0.074$ & $0.893 \pm 0.039$ \\
Polar + GT centers & $0.937 \pm 0.053$ & $0.895 \pm 0.055$ & $0.944 \pm 0.064$ & $0.937 \pm 0.040$ \\
Polar + NP centers (proposed method) & $0.932 \pm 0.027$ & $0.895 \pm 0.033$ & $0.915 \pm 0.040$ & $0.973 \pm 0.018$ \\
\hline
\end{tabular}
\caption{A summary of the mean segmentation results of our experiments. \textit{Non-polar} are the results of the U-Net trained using cartesian images. \textit{Polar + GT centers} are the results of the U-Net trained on polar images, using ground-truth connected component centers during inference, as an example of the best case possible results. \textit{Polar + NP centers} are the results when running inference on the polar model using center points obtained from the non-polar model predictions.}
\label{table:results}
\end{table*}

To perform evaluation, we use 3-fold cross-validation on the 18 scans. For each fold, we train a polar and non-polar model using the slices of 12 CT scans and run inference on the slices of the remaining 6 scans. All results presented in this section are averaged across each CT scan and then across the three folds.

\begin{figure}
\centering
\includegraphics[width=\columnwidth]{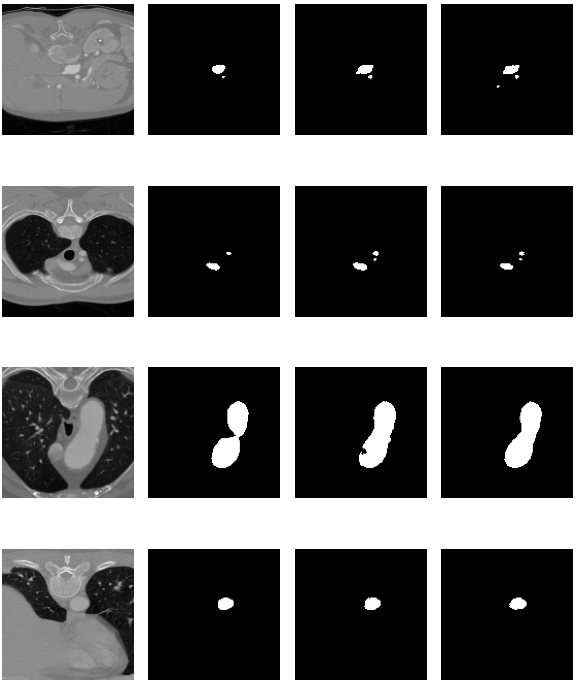}
\caption{Random examples of predictions. Columns from left to right show: the input image, the initial prediction from the non-polar network, the final fused polar prediction, the ground truth segmentation label.}
\label{fig:examples}
\end{figure}

A summary of our segmentation results is presented in Table \ref{table:results}. Random examples of segmentation results are shown in Figure \ref{fig:examples}. In the experiments in \cite{bencevicTrainingPolarImage2021} the polar networks achieve the best segmentation performance when using accurate center points during inference. As the accuracy of the center points goes down, so does the segmentation performance. The experiments in this paper follow the same pattern, the polar networks perform significantly better than the cartesian networks when using ground truth centers. However, even with less accurate centers obtained from initial rough segmentation by the non-polar network, the results yield only slightly lower Dice coefficients than when using ground-truth centers directly. The non-polar network can also be seen as a baseline model, and our approach results in a significant improvement over this baseline in all segmentation metrics.

In some problems in medical imaging, e.g. segmenting cancerous tissues, a higher recall is beneficial since the cost of missing tissues can be very high \cite{tahaMetricsEvaluating3D2015}. A key advantage of our approach is that by fusing multiple predictions and using hysteresis thresholding the threshold value can be used to impact the bias-variance tradeoff and thus increase the pixel-level recall of the segmentation. Our experiments show that the average per-patient pixel-level recall increased significantly when compared to the baseline model.

\begin{figure}
\centering
\includegraphics[width=0.8\columnwidth]{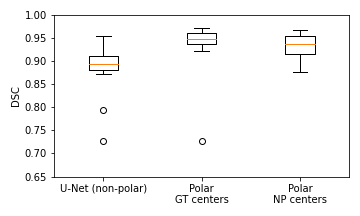}
\caption{A box plot of the per-scan Dice coefficients of our experiments. \textit{Non-polar} are the results of the U-Net trained using cartesian images. \textit{Polar + GT centers} are the results of the U-Net trained on polar images, using ground-truth connected component centers during inference. \textit{Polar + NP centers} are the results when running inference on the polar model using center points obtained from the non-polar model predictions.}
\label{fig:box}
\end{figure}

We also present the standard deviation across CT scans as a measure of segmentation reliability. Using the polar coordinates decreases the standard deviation between patients of all performance metrics, indicating that the predictions are more reliable and more robust to inter-patient differences. To further emphasize this, we present a box plot of segmentation results for each patient in Figure \ref{fig:box}. Note that, in contrast with the baseline model, when using our approach there are no outliers in the box plot.

A comparison of our results with other deep learning-based approaches for aorta segmentation in the literature is shown in Table \ref{table:comparison}. Our approach achieves performance comparable to the state of the art, and a large improvement over the baseline methods. Note that the results are evaluated on different datasets and with a different number of cases. Therefore, it is hard to objectively compare these approaches.

\begin{table}
\def\arraystretch{1.25}
\centering
\begin{tabular}{l l l c}
 \hline
 Method & DSC & mIoU & n\\ 
 \hline
Yu \textit{et al.} \cite{yuThreeDimensionalDeepConvolutional2021} & 0.958 & - & 25 \\
Fantazzini \textit{et al.} \cite{fantazzini3DAutomaticSegmentation2020} & $0.928 \pm 0.013$ & $0.866 \pm 0.023$ & 10 \\
Cheung \textit{et al.} \cite{cheungComputationallyEfficientApproach2021} & $0.912$ & - & 14 \\
Proposed method & $0.932 \pm 0.027$ & $0.895 \pm 0.033$ & 18 \\
\hline
\end{tabular}
\caption{A comparison of our approach with results reported in papers describing deep learning-based aorta segmentation methods. Note that the datasets used for obtaining the results are not the same. $n$ is the number of cases used to obtain the evaluation.}
\label{table:comparison}
\end{table}

\section{Conclusion}

In this paper, we further validate the use of polar image transformations as a tool to improve semantic segmentation performance and robustness on medical images using deep learning-based approaches. By fusing predictions of separate objects in an image we can achieve large improvements over baseline networks trained on cartesian images for segmenting the aorta. 

We show that our method can improve the segmentation performance of aorta segmentation across a variety of metrics without significantly increasing training times or the complexity of the used neural network architectures. In addition, by fusing separate predictions of different objects on the image with hysteresis thresholding we can increase pixel-level recall (at the cost of accuracy) which is often beneficial in medical image segmentation tasks.

We also show that this approach is comparable to state-of-the-art approaches for this task. This framework could be used to generally improve the performance of segmentation algorithms for various images in which multiple elliptical objects need to be segmented.

\section*{Acknowledgment}

This work was supported in part by Faculty of Electrical Engineering, Computer Science and Information Technology Osijek grant "IZIP 2022" and by the Croatian Science Foundation under Project UIP-2017-05-4968.

\bibliographystyle{ieeetr}
\bibliography{bibliography}
\end{document}